\documentclass[twocolumn,superscriptaddress,showpacs,preprintnumbers,amsmath,amssymb,floatfix]{revtex4}
\usepackage{graphicx}

\begin{document}

\title{Formation of Polymorphic Cluster Phases
for Purely Repulsive Soft Spheres}

\author{Bianca~M.~Mladek}
\affiliation{Center for Computational Materials Science and Institut f\"ur
  Theoretische Physik, Technische Universit\"at Wien, Wiedner Hauptstra{\ss}e
  8-10, A-1040 Wien, Austria}

\author{Dieter~Gottwald} 
\affiliation{Center for Computational Materials Science and Institut f\"ur
  Theoretische Physik, Technische Universit\"at Wien, Wiedner Hauptstra{\ss}e
  8-10, A-1040 Wien, Austria}

\author{Gerhard~Kahl}
\affiliation{Center for Computational Materials Science and Institut f\"ur
  Theoretische Physik, Technische Universit\"at Wien, Wiedner Hauptstra{\ss}e
  8-10, A-1040 Wien, Austria}

\author{Martin~Neumann}
\affiliation{Institut f\"ur
  Experimentalphysik, Universit\"at Wien, Strudlhofgasse 4, 
  A-1090 Wien, Austria}

\author{Christos~N.~Likos}
\affiliation{Institut f{\"u}r Theoretische Physik II,
Heinrich-Heine-Universit{\"a}t D{\"u}sseldorf,
Universit{\"a}tsstra{\ss}e 1, D-40225 D{\"u}sseldorf,
Germany}

\date{\today}

\begin{abstract}
We present results from density functional theory and computer
simulations that unambiguously predict the occurrence of first-order
freezing transitions for a large class of ultrasoft model systems into
cluster crystals. The clusters consist of fully overlapping
particles and arise without the existence of attractive forces. The
number of particles participating in a cluster scales linearly with
density, therefore the crystals feature
density-independent lattice constants. Clustering is accompanied by
polymorphic bcc-fcc transitions, with fcc being the stable phase at
high densities.
\end{abstract}


\maketitle

The distinguishing feature of soft matter systems is the vast
separation of length and time scales characterizing the extent and
motion of their constituent entities. Whereas soft matter mixtures are
typically solutions in a microscopic solvent, the solute particles are
complex macromolecular aggregates of mesoscopic spatial dimensions
\cite{russel}. The ability to control the architecture and chemical
nature of these macromolecules, combined with the flexibility in
influencing the solvent properties and the composition of the system,
gives rise to an unprecedented freedom in tuning the effective
interactions between the particles and opens up the possibility to
steer the macroscopic properties of the system \cite{russel,Lik01}.
The richness of spontaneously forming complexes in soft matter encompasses
length scales that exceed the dimensions of the individual
macromolecules. Indeed, the latter can self-organize in a variety of
ways, giving rise to so-called {\it hypermolecular} structures
\cite{Fre04} that encompass a large number of mesoscopically-sized
entities. Characteristic examples are the complex phases encountered
in ternary mixtures of oil, water, and amphiphilic surfactants or in
block copolymer blends, as well as the emergence of {\it cluster
formation} between colloidal particles, which has attracted a great
deal of attention recently \cite{segre:prl:01,sciortino:prl:04,
stradner:nature:04, mossa:langmuir:04,
bartlett:prl:05,bartlett:jpcm:05,petekidis:epl:05,zacca:05,sciortino}.
The
underlying physical mechanism that drives the emergence of
hypermolecular structures is widely believed to rest on the existence
of
competing interactions among the mesoscopic solute constituents.  For
example, the dominant mechanism that guarantees the stability of
finite clusters in colloidal
\cite{sciortino:prl:04,mossa:langmuir:04,bartlett:prl:05} or
biological \cite{stradner:nature:04} systems stems from the presence
of short-range attractions and long-range repulsions in their
effective interaction potential. Whereas the former provide the
driving force for unlimited cluster growth, the latter act as a
barrier against it. The efficiency of the barrier grows fast with
increasing cluster population, therefore further accumulation of
colloids into the clusters is brought to an end when a specific,
optimal cluster occupancy is reached \cite{Fre04, mossa:langmuir:04}.
Cluster formation is a highly topical issue in current soft matter
research, due to the large variety of cluster morphologies that form
\cite{sciortino:prl:04,bartlett:jpcm:05,zacca:05} and to the
relevance of these structures in
influencing vitrification and gelation
\cite{sciortino:prl:04,bartlett:jpcm:05,sciortino}.

In this Letter, we report on a different mechanism that gives rise to
a distinct type of cluster formation, and which does not rest on the
explicit existence of competing interactions.  Contrary to the cases
in Refs.\ \cite{sciortino:prl:04,stradner:nature:04,
mossa:langmuir:04,bartlett:prl:05}, the constituent particles we
consider are allowed to overlap and are purely repulsive.  Both
conditions are readily fulfilled for various types of polymeric
macromolecules, e.g., polymer chains \cite{ard:prl:00},
polyelectrolytes \cite{kon:jcp:04}, or dendrimers \cite{likos:jcp:04}.
Under certain, general conditions on the properties of the
Fourier transform $\tilde \Phi(q)$ of the 
interparticle potential $\Phi(r)$, 
we demonstrate that the particles form
aggregates that further organize into regular cluster crystals
with multiple site 
occupancy. We explicitly confirm the theoretical results 
by
performing extensive computer simulations on 
a specific system that
represents the entire class of effective interactions \cite{Lik01b}.

Ultrasoft effective interactions \cite{Lik01,klapp} hide many
surprises in the topology of their phase diagrams and the types of
crystal phases that arise, even for purely isotropic pair potentials
$\Phi(r)$ \cite{prl:stars,ziherl:prl,microgels:prl}. For the case in
which the $\Phi(r)$ is non-negative and bounded, a general
criterion determining the {\it topology} (but not the crystal
structures) of the phase diagram has been put forward
\cite{Lik01b}. If the Fourier transform $\tilde \Phi(q)$ is
non-negative (termed Q$^{+}$-class), then the system displays
re-entrant melting with an upper freezing temperature. If, on the
other hand, $\tilde \Phi(q)$ oscillates around zero (termed
Q$^{\pm}$-class), then a transition into an ordered cluster phase will
take place at arbitrary temperatures.  This corresponds exactly to the
ordered `clump phase' described in \cite{Kle94}.  The argument put
forward in Ref.\ \cite{Lik01b} rests on the fact that, except at
small densities and temperatures, the fluid state of the
systems at hand is extremely well described by the mean-field
approximation (MFA) $c(r) = -\beta\Phi(r)$, where $c(r)$ is the direct
correlation function and $\beta = (k_{\rm B}T)^{-1}$, with Boltzmann's
constant $k_{\rm B}$ and the absolute temperature $T$. Consequently,
the fluid structure factor $S(q)$ is given by $S(q) = [1 +
\beta\rho\tilde\Phi(q)]^{-1}$, where $\rho$ denotes the number
density. Consider now the Q$^{\pm}$-class and let $q_{*}$ be the
wavevector for which $\tilde\Phi(q)$ attains its minimum, {\it
negative} value. Then, $S(q)$ develops a real pole at $q_{*}$ along
the so-called $\lambda$-line \cite{Lik01b, Fin04, Arc04}, which
signals an instability of the uniform phase.  Evidently, the
$\lambda$-line has the shape $k_{\rm B}T_{\lambda} =
|\tilde\Phi(q_*)|\rho_{\lambda}$, i.e., it persists at all
temperatures.  
Since $q_{*}$ is set solely by the
interaction, the crystal lattice constant $a \propto q_*^{-1}$ should
be density-independent, at least on the freezing line.  Thus, the
number of particles in the elementary cell should change accordingly,
a requirement that can be fulfilled by the formation of multi-particle
aggregates (clusters).

Although the re-entrant melting scenario has been confirmed
\cite{microgels:prl,still2:97,Lan00,prestipino:05}, the clustering scenario has
received considerably less attention up to now. In this work, we
explicitly demonstrate its validity.  We perform a detailed
investigation of the generalized exponential model of index $n$
(GEM-$n$), $\Phi(r) = \varepsilon\exp[-(r/\sigma)^n]$, with $n = 4$.
It can be shown that $\Phi(r)$ belongs
to the Q$^+$-class for $n \leq 2$ and to the Q$^\pm$-class for
$n > 2$. For $n=2$, the Gaussian core model (GCM) of
Stillinger \cite{still:76} is recovered.
The GEM-4 is treated here as a representative of the
Q$^{\pm}$-class.
Suitably tailored dendrimers that have been
assembled in a computer simulation show evidence for a GEM-$n$-type
of effective interaction with $n > 2$, hence this model reflects the
behavior of realistic systems \cite{Mla05}.

\begin{figure}
\begin{center}
\includegraphics[width=5cm, clip=true, draft=false]
 {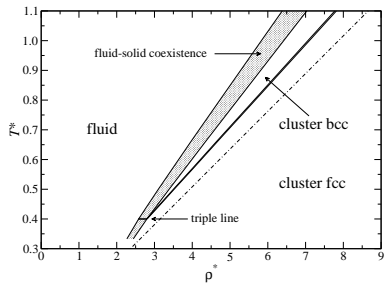}
\caption{The phase diagram of the GEM-4 model, as obtained by
  DFT. The shaded area represents the coexistence region of the liquid
  and the cluster bcc phase. The dash-dotted curve is the $\lambda$-line
  of the system, calculated in the MFA.
}
\label{phd}
\end{center}
\end{figure}

We define $\rho^{*} \equiv \rho\sigma^3$ and
$T^{*} \equiv k_{\rm B}T/\varepsilon$.
Our investigations consist of a combination of sophisticated
minimization algorithms, density functional theory (DFT), and advanced
Monte Carlo (MC) simulations. We started with a calculation at $T=0$,
allowing for the formation of clusters in which $n_c$ particles sit on
top of each other, and minimized the lattice sum with respect to the
crystal structure and $n_c$.  Since the periodic lattice that the
clusters form is {\it a priori} unknown, we took advantage of an
unbiased search strategy based on genetic algorithms \cite{Got05}.
Only fcc- and bcc-crystals were predicted, which were used as
candidates at finite temperatures. For this purpose, we employed DFT
with the highly accurate MFA-excess free energy
\cite{Lan00,ard,Arc04,acedo:santos} ${\cal F}_{\rm ex}[\rho] =
(1/2)\int {\rm d} {\bf r}_1 \int {\rm d} {\bf r}_2 \rho({\bf r}_1)
\rho({\bf r}_2) \Phi(|{\bf r}_1 - {\bf r}_2|)$.  For the one particle
density, $\rho({\bf r})$, we made the
Gaussian ansatz:
$\rho({\bf r}) \equiv \sum_{\{ {\bf R}_i \}} \rho_{\rm cl}({\bf r} - {\bf R}_i)
=
n_c (\alpha/\pi)^{3/2}\sum_{\{ {\bf R}_i \}} 
{\rm e}^{-\alpha({\bf r} - {\bf R}_i)^2}$,
where $\rho_{\rm cl}({\bf r})$ is the cluster density profile
and the vectors $\{{\bf R}_i\}$ form a Bravais lattice. 
The total free energy is
${\cal F}[\rho] = {\cal F}_{\rm id}[\rho] + {\cal F}_{\rm
ex}[\rho]$, with the ideal part
${\cal F}_{\rm id}[\rho] = k_{\rm B}T\int{\rm d}{\bf r}\rho({\bf r})
[\ln(\rho({\bf r})\Lambda^3) - 1]$, $\Lambda$ being the
thermal de Broglie wavelength. For $\alpha\sigma^2 \gtrsim 20$, $F_{\rm id}$
can be approximated analytically and 
the variational 
free energy per particle, $F/N \equiv f(n_c,\alpha)$, takes the form:
\begin{eqnarray}
\nonumber
& f(n_c, \alpha) = 
k_{\rm B}T\left[\ln n_c
+ 3\ln (\sqrt{\alpha\sigma^2/\pi}) 
\right] + 
\\
\nonumber   
& n_c\sqrt{\frac{\alpha}{8 \pi}}
\sum^{'} \int\limits_{0}^{\infty} 
{\rm d}r\,\frac{r}{R_i} 
\left[{\rm e}^{-\alpha(r - R_i)^2/2} -
{\rm e}^{-\alpha(r + R_i)^2/2} \right] \Phi(r) 
\\
& +(n_c-1) \sqrt{\frac{\alpha^3}{8 \pi}} 
\int\limits_{0}^{\infty} 
{\rm d}r\,r^2 
{\rm e}^{-\alpha r^2/2} \Phi(r), 
\label{fren:eq}
\end{eqnarray}
where the primed sum is carried over all lattice vectors
excluding ${\bf R} = 0$, $R_i = |{\bf R}_i|$, and the
irrelevant term $3k_{\rm B}T\ln(\Lambda/\sigma)$ has been dropped.
The function $f(n_c, \alpha)$ is then minimized at any state point with
respect to $n_c$ and $\alpha$.  For a given crystal
(fcc or bcc) and density $\rho^{*}$, the cluster 
population $n_c$ and the lattice vector lengths $R_i$ are
coupled to one another, $R_i/\sigma \propto l_i(n_c/\rho^{*})^{1/3}$,
where $l_i$ are lattice-specific numerical coefficients.  
At the {\it minimum} of $f(n_c,\alpha)$, 
which corresponds to a {\it mechanically} stable crystal, the 
particular property $n_c \propto \rho^{*}$ is fulfilled, 
so that the lattice constants of both the bcc- and the fcc-lattices
remain fixed at all $\rho^{*}$-values.
In particular, the nearest neighbor 
distances for bcc and fcc have the values $d_{\rm bcc} = 1.368\,\sigma$ and
$d_{\rm fcc} = 1.414\,\sigma$ 
respectively.

In order to check also the {\it thermodynamic} stability of the
crystals, we calculated the free energy of the uniform fluid employing
the MFA, $c(r) = -\beta\Phi(r)$, as a closure to the Ornstein-Zernike
relation and following the energy route to thermodynamics.
With the free energies of the
uniform and crystalline phases readily available, the phase diagram
can be drawn, which is shown in Fig.\ \ref{phd}.  At low densities,
the system is fluid. Upon compression, a first-order clustering
transition occurs, leading at sufficiently high densities to the
fcc-structure.  Above the triple temperature, $T_{\rm t}^{*} \cong
0.4$, a wedge-shaped cluster bcc-region intervenes between the fluid
and the cluster fcc.  Thus, the system also shows 
polymorphic transitions between cluster
solids. The bcc-fcc density gap is very narrow, contrary to the
fluid-crystal gap.  In agreement with the arguments in
Ref.\ \cite{Lik01b}, 
the freezing and melting curves are
almost straight lines and they precede the occurrence of the
$\lambda$-line in the phase diagram.

\begin{figure}[ht]
\begin{center}
 \includegraphics[width=8cm, clip=true, draft=false] {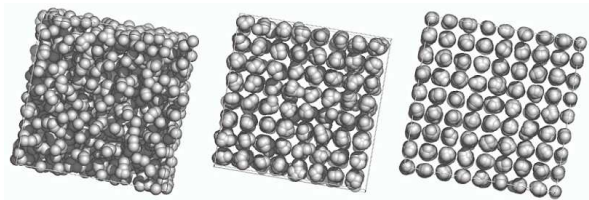}
 \caption{Three simulation snapshots of a GEM-4 system for $T^* =
 0.4$ and $\rho^* = 2.5$, 3.5, and 7 (from left to right).  Particle
 diameters are not drawn to scale but are chosen to optimize the
 visibility of the structures.}
\label{snapshot}
\end{center}
\end{figure}

To put the theoretical predictions in a stringent test, we also carried
out MC simulations in the NVT-ensemble. Typically we used systems with
approximately 5000 particles and extended the computer experiments to
$150\,000$ MC-sweeps.  Considerable speedup of the simulations was
achieved by using a discretized simulation technique \cite{Pan00}.
The symmetry of the resulting regular structures was analyzed as
proposed in Ref.\ \cite{nelson83}. 
In the simulations, we found evidence of spontaneous clustering and
crystallization, which is demonstrated by the snapshots in
Fig.\ \ref{snapshot}, all for $T^{*} = 0.4$.  Whereas in the left panel
($\rho^* = 2.5$) the system is obviously in the fluid phase, formed by
isolated particles as well as clusters, at a higher density ($\rho^* =
3.5$, middle panel), aggregates of particles have formed upon
compression. These are still `loose' and are located on a slightly
distorted fcc lattice.  At a still higher density ($\rho^* = 7.0$, right
panel), the particles are tightly bound in clusters whose
crystalline arrangement is evident.

\begin{figure}
\begin{center}
\includegraphics[width=5cm, clip=true, draft=false]
 {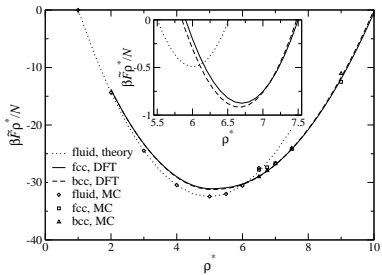}
\caption{Modified free energy density 
  $\beta\tilde F\rho^{*}/N \equiv \beta F\rho^*/N + K_1\rho^{*}$ 
  of the GEM-4 system against $\rho^*$ at
  $T^* = 1$, as obtained by DFT and MC-simulations.
  A thermodynamically irrelevant term
  $K_1\rho^{*}$ has been added for clarity of presentation.
  Inset: a close-up on the region
  of stability of the cluster bcc-lattice. Here, a different linear
  term $K_2\rho^{*}$ has been added.}
\label{energy}
\end{center}
\end{figure}

To determine the
chemical potential $\mu$ we used Widom's particle insertion \cite{Fre02}
supplemented by the overlapping distribution method
\cite{Fre02,Mla05}, and
the free energies $F_{\rm bcc}$ and $F_{\rm fcc}$ were obtained via
the Gibbs-Duhem relation.
A compilation of DFT- and MC-results for the free
energies of all phases at $T^{*} = 1$ is shown in 
Fig.\ \ref{energy}. It can be seen that the DFT-results are in
excellent agreement with simulations, a fact that amply confirms
the accuracy of the former and of the phase diagram
in Fig.\ \ref{phd}. In the inset of Fig.\ \ref{energy} we show a
close-up of the three-phase region,
to demonstrate that the cluster bcc-crystal is 
not preempted by a transition between fluid and cluster fcc.

\begin{figure}
\begin{center}
\includegraphics[width=5cm, clip=true, draft=false] 
 {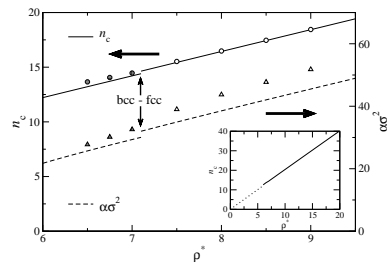}
\caption{Cluster size, $n_c$, and localization parameter, $\alpha
  \sigma^2$, for the GEM-4 system at $T^* = 1$, plotted
  against the density $\rho^*$. Lines: DFT-results; points: MC-simulations.
  There are small discontinuities at the density of the bcc-fcc transition.
  The inset shows DFT-results that corroborate 
  the $n_c \propto \rho^{*}$-relation. 
  The dashed part of the line is an extrapolation
  to low densities, for which the crystal is unstable.} 
\label{nc_alpha}
\end{center}
\end{figure}

The quantitative accuracy of DFT is not limited to 
thermodynamic quantities; it also holds for more detailed,
structural ones. In particular, we have measured in the simulations
the mean occupancy $\bar n_c$ of the clusters in the solid phases
and compared it with $n_c$ obtained from DFT. Representative
results for $T^{*} = 1$ are shown in Fig.\ \ref{nc_alpha},
where the excellent agreement between the two can be seen.
In particular, the linear dependence of $n_c$ on $\rho^{*}$
is fully confirmed. There is only a tiny difference in the
population of the bcc- and fcc-clusters, see Fig.\ \ref{nc_alpha}.
This leads to 
to a ratio $d_{\rm fcc}/d_{\rm bcc} = 1.034$, very close
to the value $\sqrt[6]{2^5}/\sqrt{3} \cong 1.029$ obtained when
$n_c$ is identical for both lattices.
At fixed $\rho^{*}$, $n_c$ is also practically
$T^{*}$-independent, so that a given crystal features a {\it single}
lattice constant in the 
whole range of its stability.

\begin{figure}
\begin{center}
\includegraphics[width=5cm, clip=true, draft=false]
 {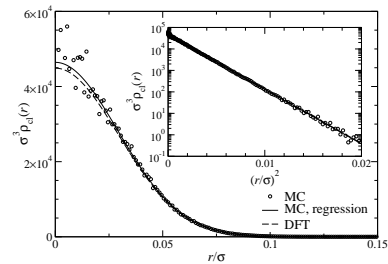}
\caption{Cluster density profile, $\rho_{\rm cl}(r)$, of the
  GEM-4 system at $T^* = 0.1$ and $\rho^* = 9$. 
  Inset: Semi-logarithmic plot of $\rho_{\rm cl}(r)$ against $r^2$,
  demonstrating the Gaussian shape of the former.}
\label{dens_prof}
\end{center}
\end{figure}

In addition, the clusters become more compact with increasing density:
this is reflected in a corresponding increase of the $\alpha$-values
in the DFT approach. Indeed, as can be seen in Fig.\ \ref{dens_prof},
the Gaussian density profile that was taken as an ansatz in the DFT
approach is in excellent agreement with the one measured in MC. This
allows for an extraction of a localization parameter $\alpha$ from the
simulation results, which can be compared to the one from DFT. The
comparison is excellent at low temperatures and worsens somewhat as
$T^{*}$ increases, however the discrepancies between the two remain
below 10\% at $T^{*} = 1.0$, as can be seen in Fig.\ \ref{nc_alpha}.
It appears that $\alpha$ also has a linear dependence on $\rho^{*}$, a
feature unknown for crystals formed by harshly-repulsive particles.
This property is intricately related to cluster formation and can be
understood as follows.  Let us fix all particles but one on their
lattice sites and consider the potential $V({\bf r})$ they exert on
this test particle,
which performs small oscillations
around a site at ${\bf r} = 0$.  Taking only the $m$ nearest
neighbor sites into account we have $V({\bf r}) = (n_c - 1)\Phi(r) +
n_c\sum\Phi(|{\bf r} - {\bf R}_{\rm nn}|)$, where the sum runs over
all nearest neighbor clusters, located at the vectors ${\bf R}_{\rm nn}$ of
length $d$.  For small $r$, $V({\bf r})$ can be expanded as $V({\bf
r}) \cong V(0) + (1/2)mn_c\Phi''(d)r^2$, since $\Phi''(0)=0$ \cite{footnote}.
Thus,
the test particle is a thermal oscillator and the density profile
$\rho({\bf r}) \propto \exp[-\beta mn_c\Phi''(d)r^2/2]$ results.  The
Gaussian shape is seen to arise from the local harmonic character of
the site potential and the localization parameter $\alpha$ can be
identified as $\alpha = \beta mn_c\Phi''(d)/2$.  Since $n_c \propto
\rho^{*}$, the linear dependence of $\alpha$ on $\rho^{*}$ follows.

All salient properties of the GEM-4-model that drive cluster formation are
common to the entire Q$^{\pm}$-class, thus the phenomena presented
here should be general. 
The spontaneous formation of clusters
appears counterintuitive at first sight as, indeed, it occurs at
the complete absence of competing interactions at the level of
the pair potential. 
The underlying reason is rather the emergence of competing
trends in the free energy, as can be seen in 
Eq.\ (\ref{fren:eq}).
The entropy loss
due to particle aggregation and the 
`self-interaction' within the cluster
(the $k_{\rm B}T\ln n_c$-term and the last term on the rhs
of Eq.\ (\ref{fren:eq}), respectively)
disfavor the growth of $n_c$. However, increasing $n_c$
also entails the gain
in avoiding close contacts with the nearest neighbors, due
to the concomitant increase of the lengths $R_i$ in the
second term on the rhs of Eq.\ (\ref{fren:eq}). 
At the same time,
excessive growth of $n_c$ is also
unfavorable, because it drives the
inter-neighbor interaction term to zero, as the
$R_i$'s then lie outside the range of $\Phi(r)$:
the entropic and self-interaction terms overtake as $n_c \to \infty$
and stop further aggregation.
Interactions with the neighbors are also indispensable for
the mechanical stability of the crystals, providing
the required restoring forces
for particle oscillations around the lattice sites.
The stability of the clusters
against both decomposition to $n_c = 1$ and unlimited growth
towards $n_c \to \infty$ is therefore seen to arise from a competition
between {\it intra}-cluster interaction and entropy, on the one hand,
and {\it inter}-cluster interaction, on the other.
The necessary 
requirements for 
this scenario
are that 
$\Phi(r)$ 
is bounded (to allow full overlaps) and
that it falls 
rapidly enough to zero for $r \to \infty$, so that $\tilde\Phi(q)$
develops oscillations. 

We have presented a detailed analysis of the
phase behavior of a particular soft-interaction-system 
representative of a broad class of effective interaction 
potentials that are realistic for ultrasoft, polymeric colloids.
A novel mechanism for the development of cluster phases has
been demonstrated to be at work, which gives rise to polymorphic
crystals with unusual structural properties. The
system is accurately described by a mean-field density functional,
as confirmed by comparison with 
computer simulations. Work along the lines of tailoring 
dendrimers that show Q$^\pm$-interactions is under way.
Future directions include the
investigation of 
anomalous
diffusion, gelation, and vitrification of such systems.

We thank H.~Fragner, D.~Frenkel, and J.~C.~P{\`a}mies for helpful discussions.
This work was supported by the \"Osterreichische Forschungsfond,
Project Nos. P15758 and P17823, by the DFG within the
SFB-TR6 and by the HPC-EUROPA project 
(RII3-CT-2003-506079).


\begin{thebibliography}{99}
\bibitem{footnote} Valid for the Q$^{\pm}$-members of the
GEM-$n$-family, $n > 2$. For $n \leq 2$, the term 
$(n_c - 1)\Phi''(0)$ renders $V''(0) < 0$ for $n_c > 1$, hence
clustering does not lead to stable crystals.
\end{thebibliography}
\end{document}